\documentstyle[12pt,twoside,fleqn,espcrc1,epsf]{article}

\newcommand{\be}{\begin{eqnarray}}
\newcommand{\ee}{\end{eqnarray}}

\newcommand{\AmS}{{\protect\the\textfont2
  A\kern-.1667em\lower.5ex\hbox{M}\kern-.125emS}}

\title{The Chiral Phase Transition}

\author{Thomas Sch\"afer\address{Institute for Nuclear Theory\\
        Department of Physics\\University of Washington\\
        Seattle, WA 98195}}

\begin{document}
\maketitle

\begin{abstract}
I review the current understanding of the chiral phase transition in QCD,
with particular emphasis on recent results in the instanton liquid model. 
\end{abstract}

\section{Introduction}

   An analysis of the spectra of particles produced in heavy ion collisions 
at CERN and AGS indicates that the excited matter created in these reactions
spends most of its life time close to the QCD phase transition \cite{Sta_96}. 
Under these circumstances, we cannot expect that the phenomena
observed in these collisions can be understood in terms of a perturbative 
plasma of quarks and gluons occupying the reaction zone. Instead, we have
to address the non-perturbative aspects of QCD associated with the phase
transition itself. Furthermore, it is these phenomena that can really 
teach us something about the structure of the QCD vacuum at zero temperature
and baryon density. In this contribution, I will try to summarize our
current theoretical understanding of the QCD phase transition. In
particular, I will report on recent progress in understanding the 
chiral phase transition in the instanton liquid model. 

\begin{figure}[t]
\begin{minipage}[b]{80mm}
\epsfxsize=7cm
\epsffile{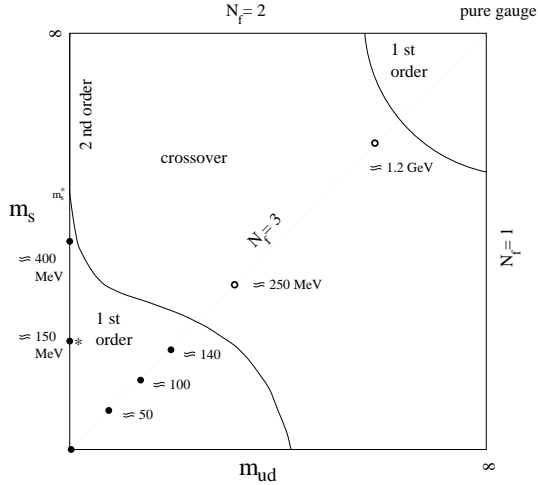}
\end{minipage}
\hspace{\fill}
\begin{minipage}[b]{75mm}
\caption{\label{fig_phase}
Schematic phase diagram for QCD in the $m_{u,d}-m_s$ plane, from
\protect\cite{IKK_96}. The points show the type of transition found
in lattice simulations with Wilson fermions. Note that QCD (*) appears
to be in the first order region, while earlier simulations with
staggered fermions placed QCD in the crossover region 
\protect\cite{BBC_90}.}
\end{minipage}
\end{figure}

   Lattice results on the QCD phase transition are discussed in 
E.~Laermann's contribution, and I will not go into detail here. 
Nevertheless, I would like to emphasize one important point. QCD
has a very rich phase structure as a function of the number of 
colors, the number of flavors and their masses. In particular, 
pure gauge QCD has a first order deconfinement transition, while 
QCD with $N_f=3$ massless flavors has a first order chiral transition, 
see figure \ref{fig_phase}. These two transitions are not connected. 
When the mass of the fermions is varied from $m=0$ to $m=\infty$ 
(corresponding to the pure gauge theory), the two first order 
transitions are separated by a region in the phase diagram in 
which there is no true phase transition, just a rapid crossover. 
Indeed, the order of the phase transition for real QCD, with two 
light and one intermediate mass flavor is still not established 
with certainty.

   The distinction between the pure gauge deconfinement and the 
light quark chiral phase transition is not just purely academic.
In fact, the two transitions have completely different energy
scales. The chiral phase transition takes place at $T_c\simeq
150$ MeV, while the purge gauge transition occurs at $T_c\simeq
260$ MeV (where the scale is set by the rho meson mass or the 
string tension). In terms
of energy density (and tax dollars!), this is an order-of-magnitude 
difference. Also, the latent heat associated with the pure gauge
transition is rather large, on the order of $1.5\,{\rm GeV}/{\rm fm}^3$,
while the chiral decondensation energy is $\Delta\epsilon\simeq
250\,{\rm MeV}/{\rm fm}^3$. This has important consequences 
for the non-perturbative gluon condensate. While most of the 
gluon condensate is removed across the pure gauge transition,
there is evidence that a significant part of it remains above the 
chiral transition.

\section{Vacuum engineering}

  In Monterey T.D. Lee reminded us that the ultimate goal of 
relativistic heavy ion collisions is vacuum engineering, the 
removal of the quark and gluon condensates present in the $T=
\mu=0$ vacuum. Here in Germany, vacuum engineering has a long 
tradition. More than three hundred years ago Otto v.~Guericke, 
after inventing a suitable pump, demonstrated the existence 
of air pressure by evacuating a pair of hollow semi-spheres. 
In QCD, we have to overcome the non-perturbative vacuum pressure 
in order to produce a perturbative vacuum state. The vacuum
pressure is determined by the the trace anomaly
\be
\label{trace_anom}
p = -\frac{1}{4}\langle T_{\mu\mu} \rangle  \;=\;
 \frac{b}{32\pi} \left\langle\alpha_s G^2
 \right\rangle - \frac{1}{4} \sum_f m_f\langle \bar q_fq_f\rangle,
\ee
where $b=11N_c/3-2N_f/3$ is the first coefficient of the beta
function. Using the canonical values of the condensates, this
relation gives $p=500\,{\rm MeV}/{\rm fm}^3$. At low temperature,
the $T$-dependence of the condensates is determined by chiral
perturbation theory \cite{Leu_96}
\be
\langle\bar qq\rangle &=& \langle\bar qq\rangle_0 
 \left\{ 1- \frac{N_f^2-1}{3N_f}\left(\frac{T^2}{4f_\pi^2}\right) 
 - \frac{N_f^2-1}{18N_f^2} \left(\frac{T^2}{4f_\pi^2}\right)^2 
 + \ldots \right\}, \\
 \langle \alpha_s G^2\rangle &=& 
 \langle \alpha_s G^2\rangle_0
 -\frac{4\pi^4}{405b}N_f^2(N_f^2-1)\frac{T^8}{f_\pi^4}
 \left(\log\left(\frac{\Lambda}{T}\right)-\frac{1}{4}\right)
 -\ldots\; . 
\ee
To leading order, the $T$-dependence of the quark condensate
has a very simple interpretation in terms of the number of 
thermal pions times the pion matrix element of the quark
condensate, $\langle \pi|\bar qq|\pi \rangle = -\langle\bar 
qq\rangle/f_\pi^2$ \cite{GL_87}. This means that pions
act as a vacuum cleaner for the chiral condensate, each thermal
pion removes $\sim 5$ $\bar qq$ pairs. On the other hand, the 
gluon content of the pion is rather small. Naively extrapolating 
these results to larger temperatures, one expects chiral 
symmetry restoration to occur at $T\simeq 260$ MeV, while 
the gluon condensate is essentially $T$-independent. This
estimate is not strongly affected by higher loop corrections
in ChPTh. Clearly, chiral perturbation theory was not meant 
to be used near the transition. Nevertheless, the result 
indicates that something more than just pions is needed
to restore the symmetry at the expected temperature. It 
also shows that even if the chiral expansion is apparently
convergent, the neglected (exponentially small) terms need
not be small. 

   The thermodynamics of the phase transition is usually
described in terms of a bag model equation of state. This 
EOS directly incorporates the idea that
the transition takes place as soon as the perturbative 
pressure from quarks and gluons can overcome the 
non-perturbative bag pressure in the QCD vacuum. For 
$N_f=2$ and $B=500\,{\rm MeV}/{fm}^3$ from (\ref{trace_anom})
this gives the estimate $T_c=[(90B)/(37\pi^2)]^{1/4}\simeq 
180$ MeV. Analyzing lattice thermodynamics in more 
detail one finds that this estimate is too large, because
only $\sim 1/2$ of the bag pressure is removed across
the phase transition \cite{Den_89,AHZ_91,KB_93}.

   An even simpler approach to estimate the critical 
temperature is based on the idea that the transition 
occurs when thermal hadrons begin to overlap. The density 
of hadrons becomes of order $1\,{\rm fm}^{-3}$ near 
$T\simeq 200$ MeV. This is reasonably close to $T_c$
(although, if $T_c$ is really 150 MeV then $n_{had}
(T_c)\simeq 0.15\,{\rm fm}^{-3}$). Nevertheless, this 
kind of argument is not correct in general. A good 
example is pure gauge QCD, where $T_c \simeq 250$ MeV, 
but the lightest state is at $\sim 1.7$ GeV, so the 
particle density below $T_c$ is small, on the order of
$n\simeq 0.005\,{\rm fm}^{-3}$. The lesson is that a 
first order transition occurs if the high $T$ (QGP) and
low $T$ (hadronic) phases have the same free energy; the
phase transition point cannot be inferred from looking
at the low temperature phase only.  
   
\section{QCD near $T_c$}

  Chiral perturbation theory is based on a non-linear 
effective lagrangian in which the $\sigma$, the chiral
partner of the pion, is eliminated. Near $T_c$, this 
description is not expected to be useful. However, in the vicinity
of a second order phase transition, universality implies that 
critical phenomena are governed by an effective Landau-Ginzburg 
action for the chiral order parameter. In the case of QCD 
with two (massless) flavors the order parameter is a four-vector
$\phi^a=(\sigma,\vec\pi)$. Universality makes definite 
predictions for the critical behavior of $\langle\bar qq\rangle$,
the chiral susceptibility and the specific heat near $T_c$
\cite{PW_84,Wil_92,RW_93}. At the moment, these predictions 
appear to be in agreement with lattice gauge results \cite{KL_94}, 
but the issue has not been completely settled \cite{KK_95}. 

  I would like to make a few comments concerning the role 
of universality arguments. First, it is important to clearly
distinguish between the low energy chiral lagrangian (or the 
linear $\sigma$-model used as an effective lagrangian, see e.g.
\cite{BK_96}) and the 
effective action for the order parameter near $T_c$. The 
Landau-Ginzburg action is a three dimensional action for
static modes only. It is applicable only near $T_c$. In 
particular, the parameters in the effective action,
the $\pi,\sigma$ masses and couplings, are completely 
independent of the parameters used in the linear $\sigma$-model 
at $T=0$. My second point concerns the thermodynamics
of the phase transition. The effective action describes the
singular part of the free energy only. In QCD we expect a
large change in the free energy that corresponds to the 
release of 37 (quark and gluon) degrees of freedom. This
means that in practice, the non-universal, regular, part
of the free energy will most likely dominate the universal,
singular, contribution. 

   Universality predicts the behavior of three dimensional
(screening) correlation functions near $T_c$. The corresponding
screening masses have also been studied in some detail on the
lattice, see section 5. In practice, however, we are more 
interested in dynamical (temporal) masses, corresponding to
poles of the spectral function in energy, not momentum. These
quantities are hard to extract on the lattice, although some
exploratory studies have been made \cite{BGK_94}. In addition to 
that, we have made significant progress in studying temporal correlation 
functions in the instanton liquid model (see below). The 
only general approach to the problem that we have available
at the moment are QCD sum rules.

  The general strategy is easily explained. It is based on matching 
phenomenological information contained in hadronic spectral function
with perturbative QCD, using the operator product expansion. At
finite temperature, the sum rules are of the type
\be
 c_0 \log\omega^2 +c_1\langle O_1\rangle \frac{1}{\omega^4}
 + c_2\langle O_2\rangle \frac{1}{\omega^6} +\ldots =
 \int du^2\frac{\rho(u^2)}{u^2-\omega^2},
\ee
where $\rho(\omega^2)$ is the spectral function at $\vec p=0$, $c_i$
are temperature independent coefficients that can be calculated in
perturbative QCD and $\langle O_i\rangle$ are temperature dependent
condensates. If there is a range in energies in which both the OPE
has reasonable accuracy and the spectral representation is dominated
by the ground state, we can use the sum rules to make predictions
about ground state properties.

  In practice, this is a difficult game, even at zero temperature.
At $T\neq 0$, additional problems arise because we do not know the
$T$-dependence of the condensates and there is little phenomenological
guidance concerning the form of the spectral functions. For this 
reason, reliable predictions can only be made at small temperature.
The most systematic studies have been made in the vector meson 
channels $\rho$ and $a_1$ \cite{EI_93,EI_95,HKL_93}. To order $T^2$, 
there is no shift in the resonance masses. The only effect is mixing 
between the $\rho$ and $a_1$ channels, which is caused by scattering 
off thermal pions. At order $T^4$, masses start to drop. It is
interesting to note that at this order, the mass shift is not
controlled by the quark condensate, but the energy momentum tensor
in a thermal pion gas. 

\section{The instanton liquid at finite temperature}

  In order to make progress we need a more detailed picture of  
the chiral phase transition. In particular, we would like to
understand the mechanism of the transition and the behavior of
the condensates and hadronic correlation functions in the  
transition region. In the following, I will argue that 
important progress in this direction has been made in the
context of the instanton liquid model. In this model, 
chiral symmetry breaking is caused by the delocalization 
of quark zero modes associated with instantons. Chiral
restoration takes place when instantons and antiinstantons
form molecules, the quark modes become localized and the
quark condensate is zero. 

\begin{figure}[t]
\epsfxsize=14cm
\epsffile{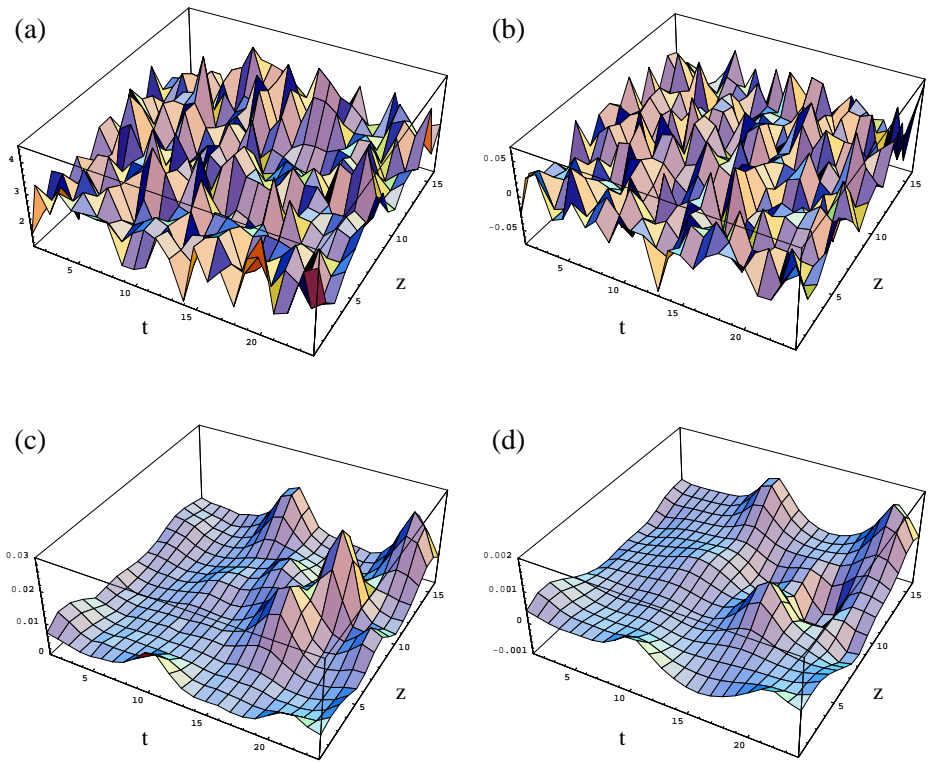}
\vspace*{-0.5cm}
\caption{\label{fig_cool}
Instanton content of a typical $T=0$ gauge configuration, from 
\protect\cite{CGHN_94}. Figs. (a) and (c) show the field strength
while (b) and (d) show the topological charge density. The upper 
panel shows the original configuration and the lower panel the 
same configuration after 25 cooling sweeps.}
\end{figure}

   The essential assumption underlying the instanton model
is that the (euclidean) QCD partition function
\be
Z= \int DA_\mu \exp(-S)\prod_f^{N_f} \det(iD\!\!\!\!/\,+im_f) ,
\ee
is dominated by classical gauge configurations called instantons. 
Instantons describe tunneling events between degenerate vacua.
As usual, tunneling lowers the ground state energy. This is 
why instantons contribute to the vacuum energy density and
pressure in the QCD vacuum. The instanton solution is 
characterized by 12 parameters, position (4), color orientation 
(7) and size (1). An ensemble of interacting instantons is 
described by the partition function 
\be
\label{Z}
Z= \sum_{N_+ N_-} {1 \over N_+ ! N_- !}\int
   \prod_i^{N_+ + N_-} [d^4z_idU_id\rho_i\; d(\rho_i)]
   \exp(-S_{int})\prod_f^{N_f} \det(iD\!\!\!\!/\,+im_f) \; ,
\ee
where $N_+$ and $N_-$ are the numbers of instantons and antiinstantons and 
$d(\rho)$ is the semi-classical instanton density calculated by 't Hooft. 
There are two important pieces of evidence that suggest that instantons 
play an important role in the QCD vacuum. One is provided by extensive 
calculations of hadronic correlation functions in the instanton
liquid model \cite{SV_93b,SSV_94,SS_96b}. These correlators agree both 
with phenomenological information \cite{Shu_93} and lattice calculations 
\cite{CGHN_93b}. The second comes from direct studies of instantons 
on the lattice. An example is shown in figure \ref{fig_cool}. Using 
a procedure called ``cooling" one can relax any given gauge field 
configuration to the closest classical component of the QCD vacuum. 
These configurations were found to be ensembles of instantons and 
antiinstantons. The MIT group concludes that the instanton density 
is $(N/V)\simeq (1.4-1.6)\,{\rm fm}^{-4}$ while  the typical size 
is about $\rho\simeq 0.35$ fm \cite{CGHN_94}. These numbers are in 
good agreement 
with the instanton liquid parameters $(N/V)= 1\,{\rm fm}^{-4}$, 
$\rho=1/3\,{\rm fm}$ proposed by Shuryak a long time ago \cite{Shu_82a}. 
What is even more important is that hadronic correlation functions 
remain practically unchanged during the cooling process. This 
suggests that instantons play a dominant role in generating the 
spectrum of light hadrons. 

\begin{figure}[t]
\begin{minipage}[b]{80mm}
\epsfxsize=7cm
\epsffile{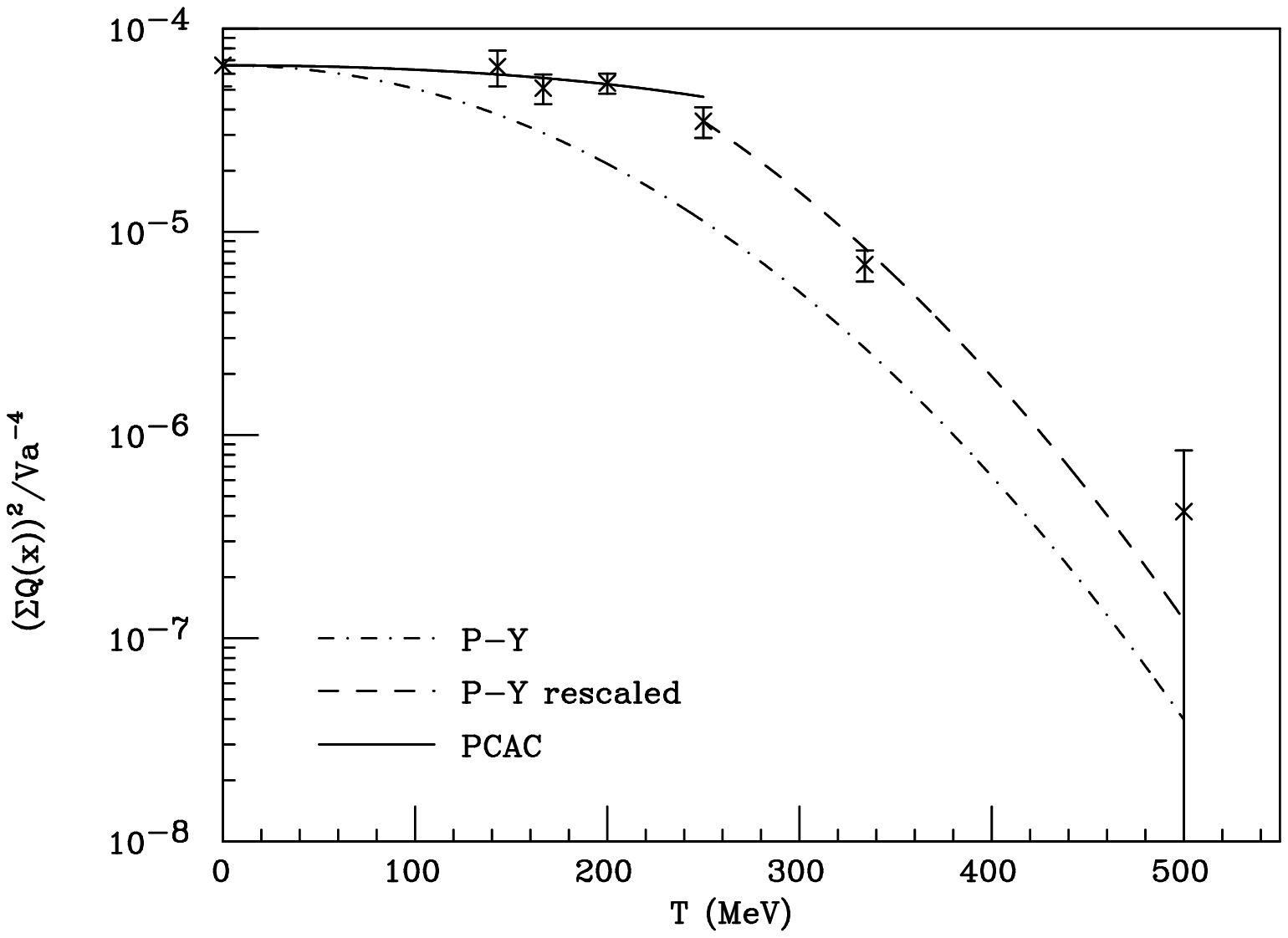}
\vspace*{-1.0cm}
\caption{\label{fig_chu}
$\chi_{top}$ as a function of temperature in quenched
QCD, from \protect\cite{CS_95}.}
\vspace*{0.5cm}
\epsfxsize=7cm
\epsffile{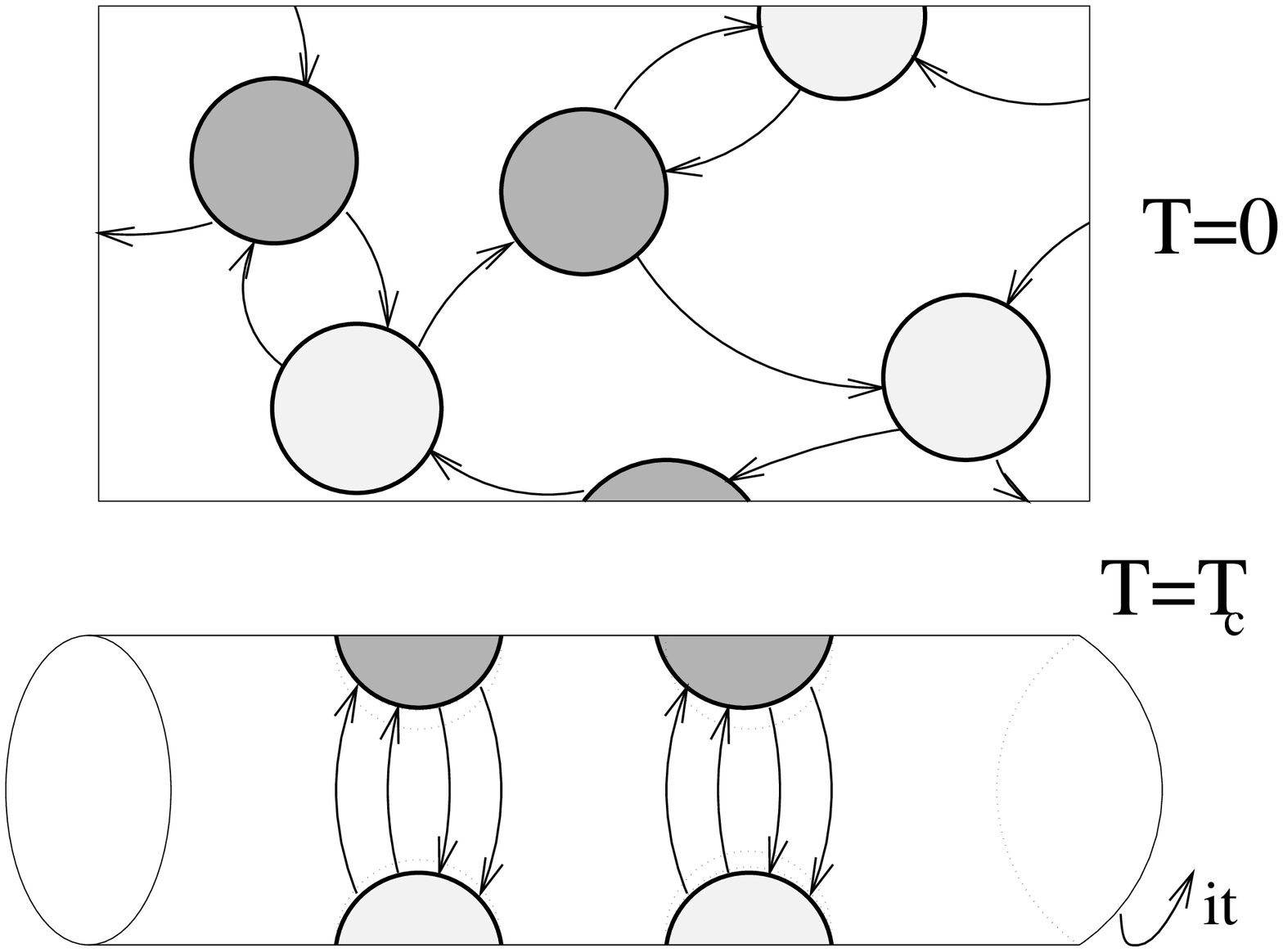}
\caption{\label{fig_schem}
Schematic picture of the phase transition in the interacting instanton
liquid.}
\end{minipage}
\hspace{\fill}
\begin{minipage}[b]{75mm}
\epsfxsize=7cm
\epsffile{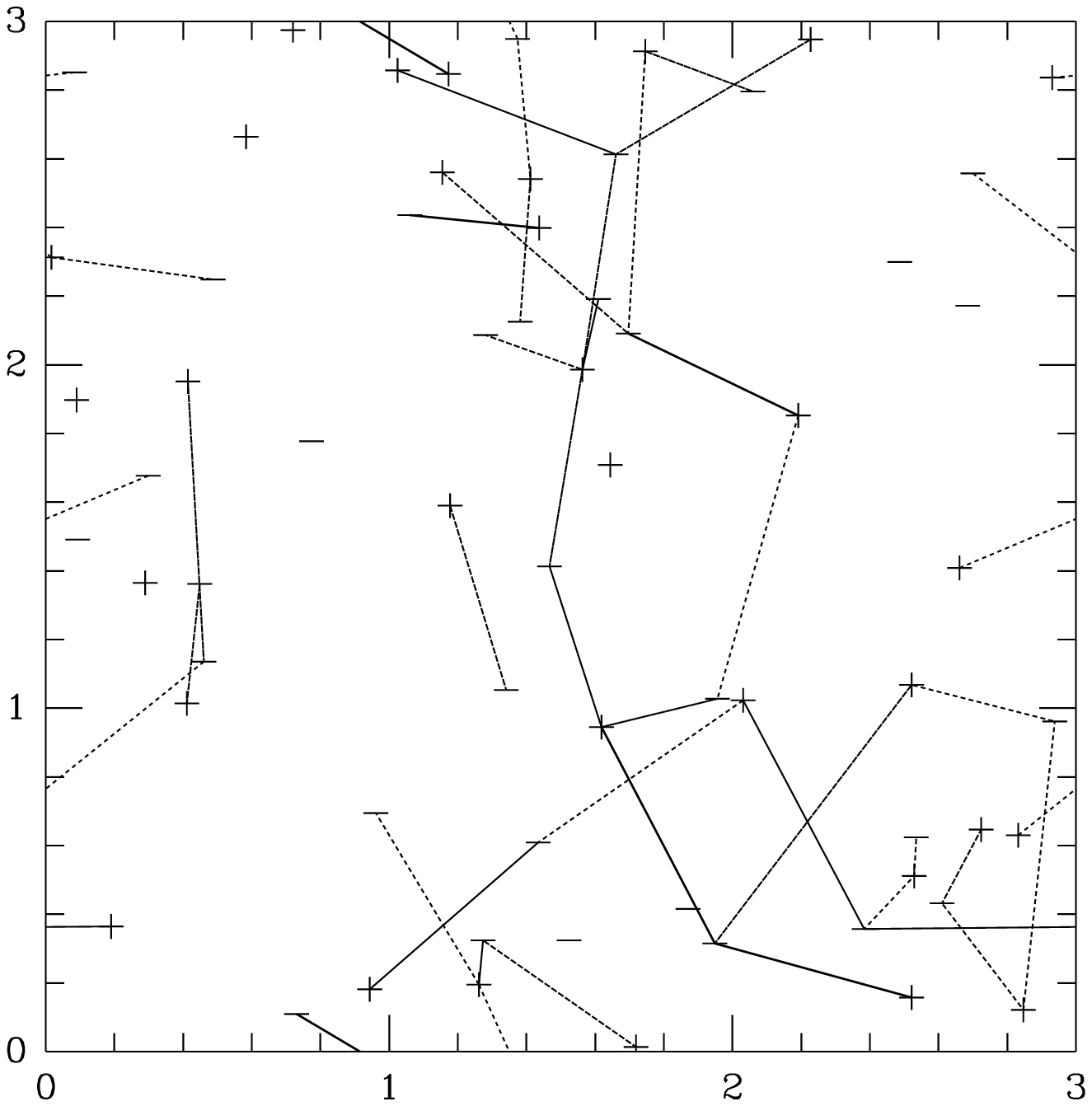}
\vspace*{1.5cm}
\epsfxsize=7cm
\epsffile{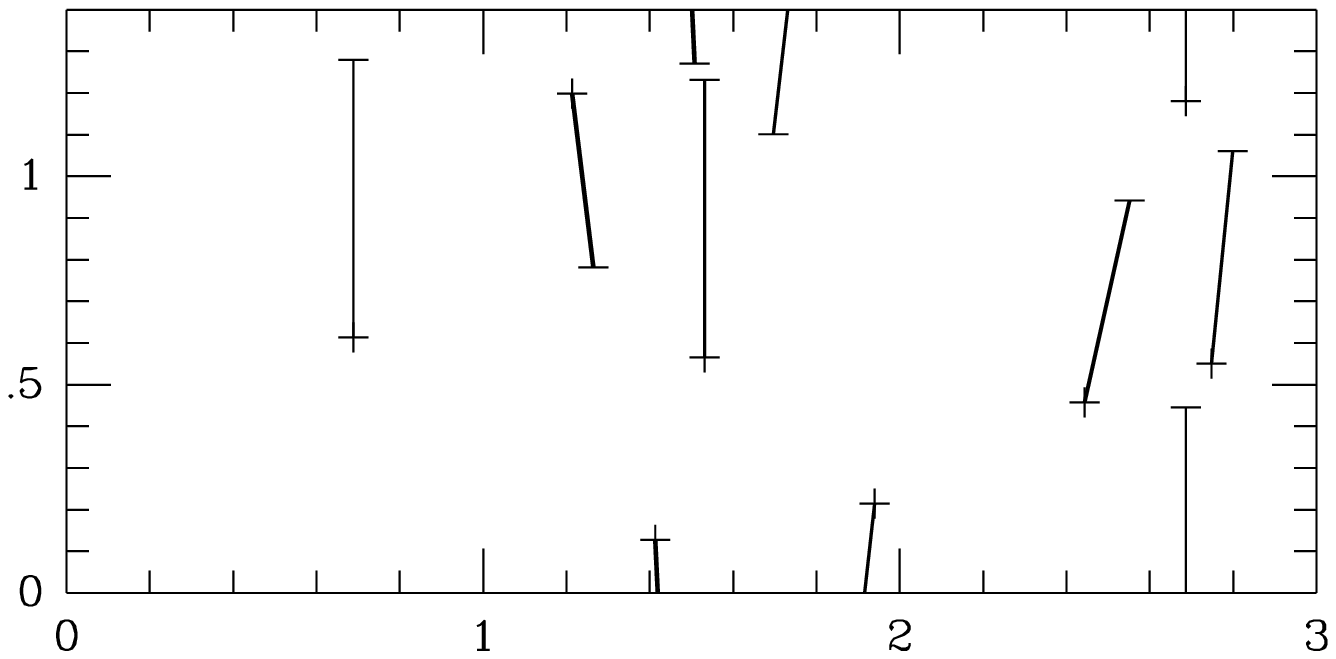}
\caption{\label{fig_configs}
Instanton configurations below and above the phase transition.}
\end{minipage}
\end{figure}

  Recently, the role of instantons at finite temperature has 
been reevaluated. The semi-classical expression for the instanton
density at finite temperature contains a suppression factor $\sim\exp(
-(2N_c/3+N_f/3)(\pi\rho T)^2)$ \cite{GPY_81} which comes mostly from 
Debye screening of the instanton field. From this expression we would 
expect that instantons are significantly suppressed ($\sim 0.2$) near 
$T_c$. This is not correct. The perturbative expression for the 
Debye mass is not applicable until the temperature is significantly
above $T_c$. We have already seen that the gluon condensate is 
essentially temperature independent at small $T$. A similar 
calculation can also be performed for the instanton density,
with the same conclusion \cite{SV_94}. This result was confirmed
in a number of lattice calculations (in quenched QCD) 
\cite{CS_95,IMM_95,ADD_96}, see figure \ref{fig_chu}. 
The figure shows the topological susceptibility which,
in quenched QCD, is roughly equal to the instanton density.
The Pisarski-Yaffe prediction (labeled P-Y) clearly underpredicts
the topological susceptibility near $T_c$. The dashed curve which 
fits the data above $T_c$ corresponds to the PY-prediction with 
a shifted temperature $T\to(T-T_c)$. 

  If instantons are not suppressed around $T_c$, then the chiral
phase transition has to be caused by a rearrangement of the 
instanton liquid. A mechanism for such a rearrangement, the 
formation of polarized instanton-antiinstanton molecules, was 
proposed in \cite{IS_94,SSV_95}. In the presence of light fermions, 
instantons interact 
via the exchange of $2N_f$ quarks (see fig. \ref{fig_schem}). 
The amount of correlations in the instanton ensemble depends on the
competition between maximum entropy, which favors randomness, and
minimum action, which favors the formation of instanton anti-instanton
pairs. At low temperatures the instanton system is random and chiral
symmetry is broken. At high temperatures the interaction in the 
spacelike direction becomes screened, whereas the periodicity of the
fields in the timelike interaction causes the interaction in that 
direction to be enhanced. Schematically, the fermion determinant for
one pair looks like
\be
\det(iD\!\!\!\!/\,)\sim |\sin(\pi T \tau)/\cosh(\pi T r)|^{2N_f},
\ee
where $\tau$ and $r$ are the separations in the temporal and spatial
direction. This interaction is maximal for $r=0$ and $\tau=\beta/2=
(1/2T)$ which is the most symmetric configuration of the $I\bar I$
pair on the Matsubara torus. 

  In numerical simulations, we find the transition to a correlated 
system in which chiral symmetry is restored at $T_c\simeq 130$ MeV
\cite{SS_96}. Typical instanton configurations below and above $T_c$ 
are shown in figure \ref{fig_configs}. The plots are projections of a 
four dimensional box into the $x\tau$-plane. At high temperature, the
imaginary time axis is short. The location of instantons and 
antiinstantons is denoted by $\pm$ signs, while the lines connecting
them indicate the strength of the fermionic ``hopping" matrix
elements. Below $T_c$, there is no clear pattern. Instantons 
are either isolated or part of larger clusters. Following the
hopping matrix elements, quarks can propagate over large distances
and form a condensate. Above $T_c$, instantons are bound into pairs. 
The propagation of quarks in the spatial direction is suppressed
and no condensate is formed. 

   More details are provided in figure \ref{fig_dens}. The quark
condensate is practically $T$-independent at small temperature,
but drops rapidly above $T=100$ MeV. At small $T$, the instanton
density rises slightly\footnote{This might very well be an artefact.
The $T=0$ point is not a true zero temperature calculation.}. It
drops above $T=100$ MeV, but retains about half of its $T=0$
value near $T_c$. Similarly, the instanton related free 
energy\footnote{Figure \protect\ref{fig_dens}b only shows 
the instanton-related part of the free energy, the full
free energy $F=-p$ has to be a monotonically decreasing function
of $T$.} does not vanish at $T_c$. Instantons still contribute 
to the energy density and pressure above $T_c$. 

   It is interesting to study the phase diagram in more detail.
For realistic QCD, the transition appears to be weakly first 
order (the discontinuity in the free energy is smoothed out 
in a finite volume and cannot be seen in figure \ref{fig_dens}).
In the case of two flavors, we see a second order phase transition
with critical exponents consistent with the $O(4)$ universality
class. For more flavors, the transition temperature drops until 
(around $N_f=5$ massless flavors) chiral symmetry is restored
in the ground state even at $T=0$.

\section{Hot hadrons}

\begin{figure}[t]
\begin{minipage}[b]{80mm}
\epsfxsize=7cm
\epsffile{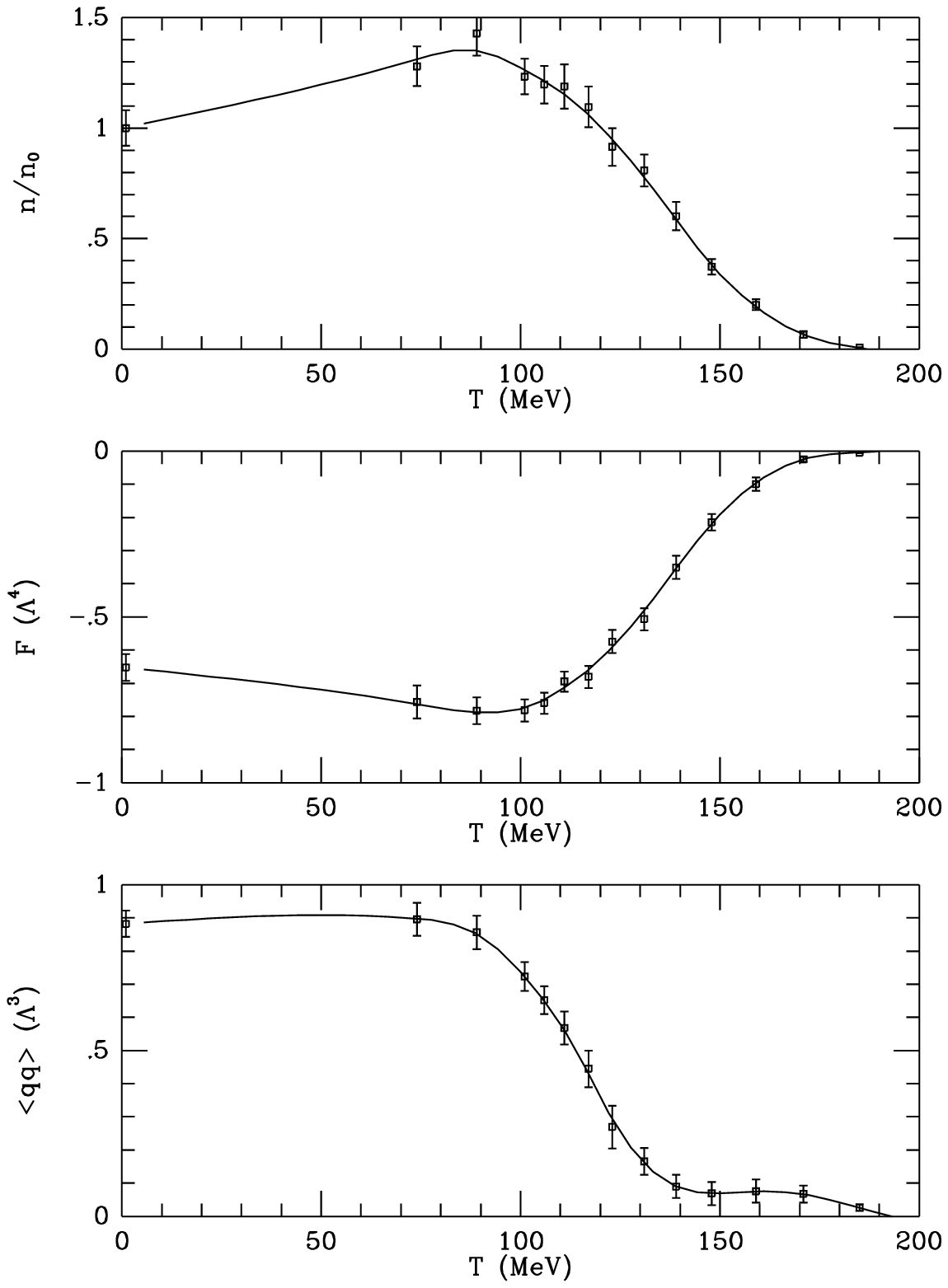}
\caption{\label{fig_dens}
Instanton density, free energy and quark condensate as a
function of $T$.}
\end{minipage}
\hspace{\fill}
\begin{minipage}[b]{75mm}
\epsfxsize=7cm
\epsffile{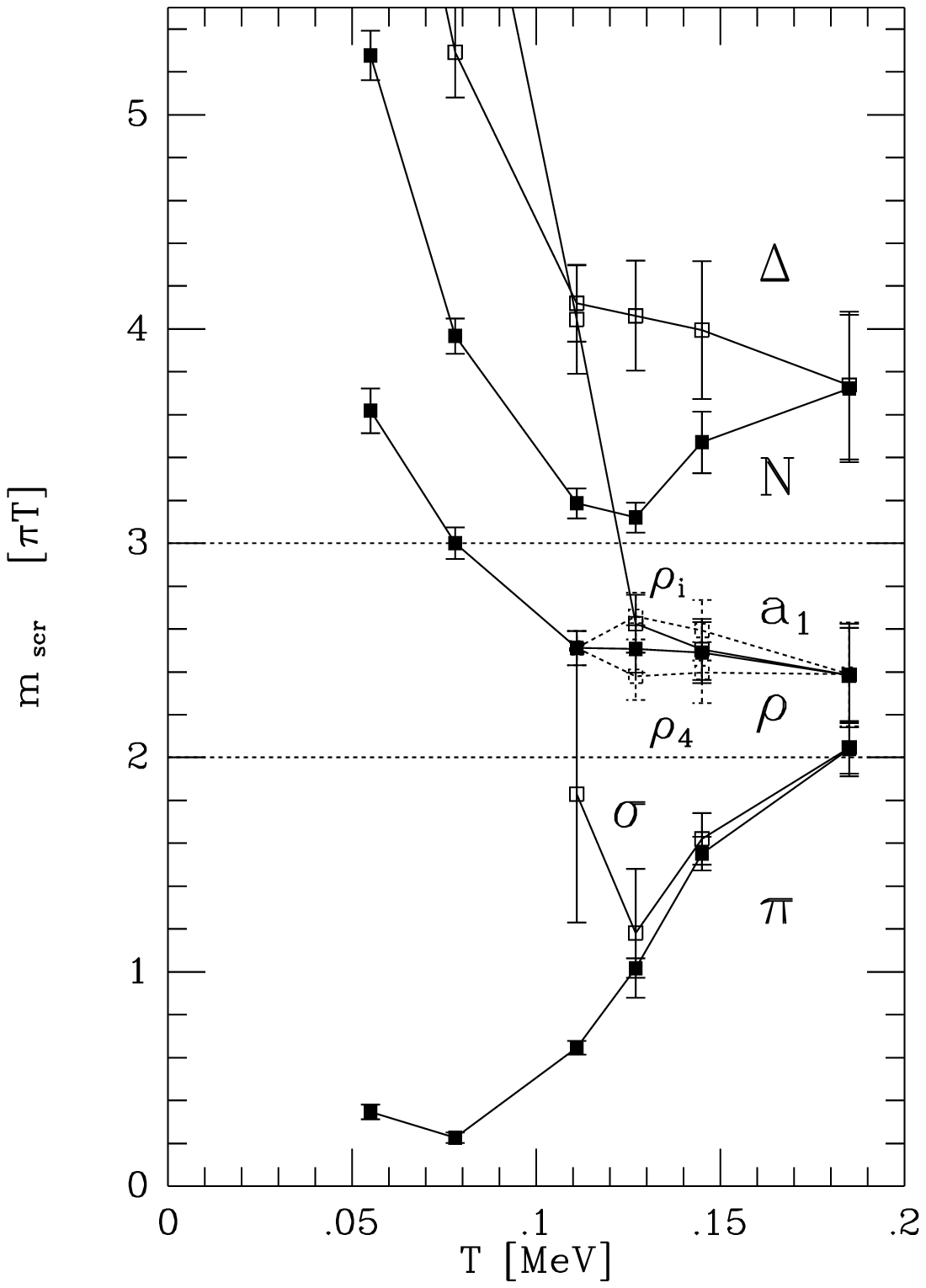}
\caption{\label{fig_scr}
Screening masses in the instanton liquid model.}
\end{minipage}
\end{figure}

  We have studied both temporal (related to the spectral function
in energy) and spatial (related to screening masses) correlation
functions at finite temperature \cite{SS_96b}. Figure \ref{fig_scr}
shows the spectrum of spacelike screening masses. First, we clearly
observe that chiral symmetry is restored at $T\simeq 130$ MeV. 
Chiral partners, like the $(\pi,\sigma)$ and $(\rho,a_1)$ become 
degenerate at $T_c$. Second, even above $T_c$, the scalars $\pi$
and $\sigma$ are significantly lighter than the vectors $\rho$
and $a_1$. This result is in agreement with lattice calculations
\cite{TK_87,Goc_91}. The spectrum of screening masses is also
in qualitative agreement with the predictions from dimensional
reduction (DR) \cite{EI_88,HZ_92,KSB_92}, but DR fails to account
for the attraction seen in the scalar channels. 

\begin{figure}[t]
\epsfxsize=14cm
\epsffile{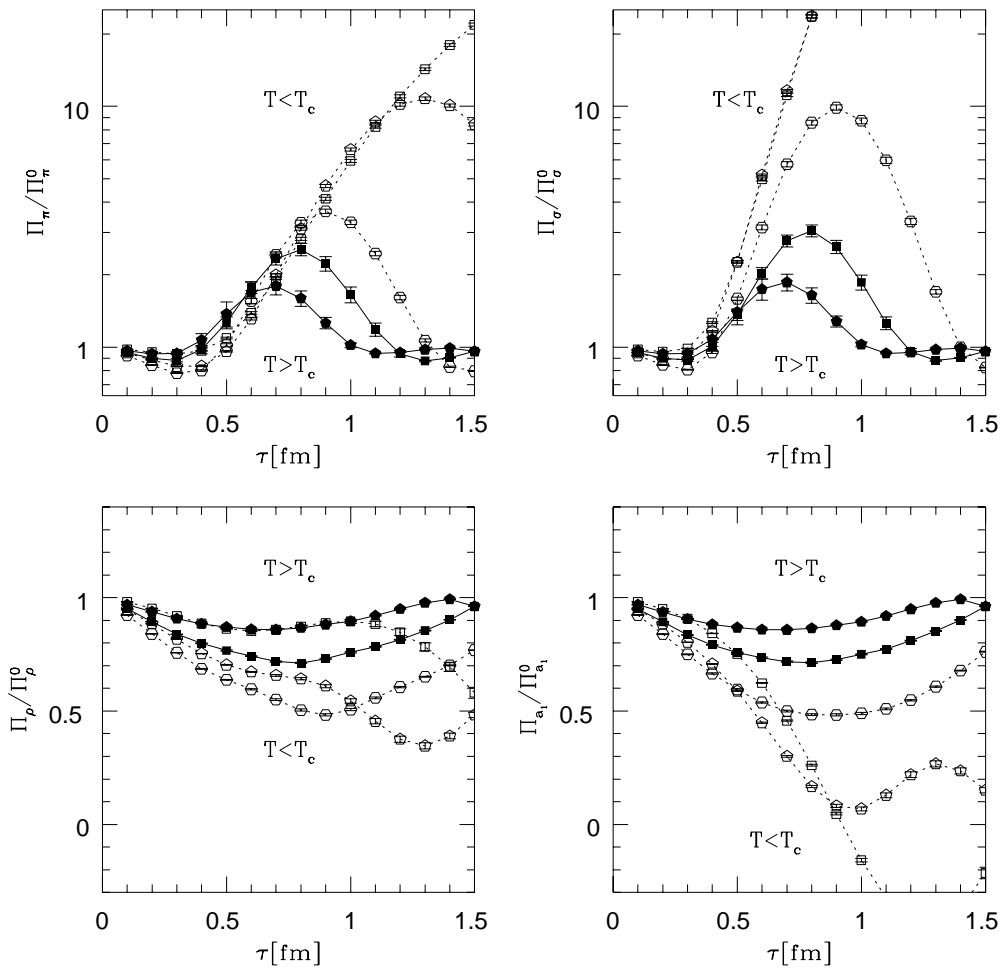}
\vspace*{-1cm}
\caption{\label{fig_tempcor}
Temporal correlation functions in the instanton liquid model.}
\end{figure}

  More interesting is the behavior of temporal correlation functions,
shown in figure \ref{fig_tempcor}. The correlation functions are 
normalized to free quark propagation at the same temperature, so 
all correlators start at 1 for $\tau=0$. Below $T_c$, the data 
points are denoted by open squares ($T=0$), pentagons, etc., 
while the high temperature points are denoted by solid squares and
pentagons. Above $T_c$, the period of the correlation functions 
becomes short, $\beta=T_c^{-1}\sim 1.5$ fm. This means that 
all the spectral information is contained in a very short interval
$0<\tau<\beta/2\sim 0.7$ fm. This fact makes it very difficult 
to extract thermal masses from the correlation functions (or 
to decide whether there are narrow poles in the spectral function
at all). 

  Again, we observe that above $T_c$ the correlation functions 
of chiral partners become equal. What is more remarkable is the
fact that in the regime $\tau<(\beta/2)$, which is not (directly)
affected by the periodic boundary conditions, the pion\footnote{
The $\sigma$ correlation is even larger than the $\pi$ below $T_c$ 
because it receives a disconnected contribution from the quark 
condensate.} correlation function is almost as large as it is at
$T=0$. This suggest that there is still a $(\pi,\sigma)$-mode above 
$T_c$ \cite{HK_85,SS_95b}. No such effect is seen in the vector 
channels. The small resonance contribution seen in the $\rho$ 
channel at small temperature quickly melts and above $T_c$ the 
correlation function is consistent with the propagation of
two independent quarks with a small residual chiral mass (a mass 
term in the vector part of the propagator that does not violate 
chiral symmetry). 

  What effect causes the resonance behavior seen in the scalar 
channels? At zero temperature, the instanton induced interaction 
between quarks is conveniently discussed in terms of an effective 
four fermion interaction\footnote{Strictly speaking, a flavor 
antisymmetric $2N_f$ fermion interaction. However, for $N_f=3$
and broken chiral symmetry (either spontaneous or explicit), we 
can absorb two zero modes into the instanton density.} 
\cite{tHo_76b,SVZ_80b}
\be
\label{leff}
  {\cal L}=  \int d(\rho)d\rho\, \frac{(2\pi\rho)^4}{8(N_c^2-1)} \left\{
     \frac{2N_c-1}{2N_c}\left[ 
     (\bar\psi\tau^{-}_a\psi)^2 + (\bar\psi\tau^{-}_a\gamma_5\psi)^2 
     \right]
     - \frac{1}{4N_c} (\bar\psi\tau^{-}_a\sigma_{\mu\nu}\psi)^2 \right\} ,
\ee 
where $d(\rho)$ denotes the density of instantons. Here, $\psi$ is an
isodoublet of quark fields and the  four vector $\tau^{-}_a$ has  
components $(\vec\tau,i)$ with $\vec\tau$ equal to the Pauli
matrices acting in isospace. The interaction (\ref{leff}) successfully  
explains  many properties of the ($T$=0) QCD correlation functions, most 
importantly the strong attraction seen in the pion channel.

  The effective lagrangian (\ref{leff}) comes from the $2N_f$ zero 
modes associated with an individual instanton. It is derived under 
the assumptions that instantons are sufficiently dilute and 
completely uncorrelated. Above $T_c$, the collective coordinates 
of instantons and antiinstantons are no longer random, but become 
strongly correlated. The four fermion interaction induced by 
polarized instanton-antiinstanton molecules is given by
\cite{SSV_95}
\be
\label{lmol}
 {\cal L}_{mol\,sym}&=& G 
        \left\{ \frac{2}{N_c^2}\left[ 
        (\bar\psi\tau^a\psi)^2-(\bar\psi\tau^a\gamma_5\psi)^2 
        \right]\right. \nonumber \\ 
         & & - \;\,\frac{1}{2N_c^2}\left. \left[
        (\bar\psi\tau^a\gamma_\mu\psi)^2+(\bar\psi\tau^a\gamma_\mu\gamma_5
        \psi)^2 \right] + \frac{2}{N_c^2}
        (\bar\psi\gamma_\mu\gamma_5\psi)^2 \right\} + {\cal L}_8,
\ee
where the coupling constant $G$ is determined by the number of 
correlated pairs (and their overlap matrix element) and ${\cal L}_8$ 
is the color octet part of the interaction. $G$ is not strong enough
in order to cause quarks to condense. Nevertheless, molecules produce 
a significant attractive interaction in the $\pi$ and $\sigma$ channels.  

  A problem that has received a lot of attention recently 
is the fate of the $U(1)_A$ anomaly at finite temperature
\cite{Shu_94,KKL_96,HW_96,Coh_96,LH_96,EHS_96}. Given the
large $\eta'-\pi$ splitting any tendency towards $U(1)_A$
could lead to rather dramatic effects in heavy ion collisions. 
Two experimental signatures that have been discussed are the 
$\eta/\pi$ ratio \cite{HW_96} measured by the WA80 collaboration
\cite{WA80_94} and the possibility that the $\eta'$ Dalitz
decay \cite{KKL_96} contributes to the enhancement in low mass 
dileptons seen by CERES \cite{CER_95c}.

  The anomaly is related to the presence of zero modes in the 
spectrum of the Dirac operator. Instantons can absorb $N_f$
left handed quarks of different flavors and turn them into
right handed quarks, violating axial charge by $2N_f$ units. 
This is the process described by the 't Hooft vertex (\ref{leff}).
Inserting the 't Hooft interaction into the $\eta'$ correlation
function splits the $\eta'$ from the pion\footnote{In QCD, flavor
symmetry is broken and part of the splitting comes from the 
strange quark mass. However, in the absence of the anomaly we
would expect the $\eta-\eta'$ mixing to be almost ideal (similar
to the $\rho-\phi$ system). In this case, there is a non-strange
$\eta$ which is almost degenerate with the $\pi$.}. Above $T_c$
isolated instantons disappear and instanton-antiinstanton molecules
do not violate $U(1)_A$. However, the 't Hooft operator can induce 
a tunneling event (instanton) all by itself. This can be seen as 
follows. If we keep a small current quark mass, the density of 
isolated instantons above $T_c$ is proportional to $m^{N_f}$. When 
we calculate a $U(1)_A$ violating observable, there are $N_f$
propagators\footnote{In the (academic) case of three massless
flavors, $U_A(1)$ violation does not affect the $\eta'$ 
correlation function above $T_c$ because the third quark 
in the 't Hooft vertex cannot be absorbed. In real QCD,
the strange quark can be absorbed by a mass insertion.}, 
each of which has a zero mode contributing a factor $1/m_f$. 
As a result, $U(1)_A$ is broken in the chiral limit $m_f\to 0$. 

\begin{figure}[t]
\epsfxsize=14cm
\epsffile{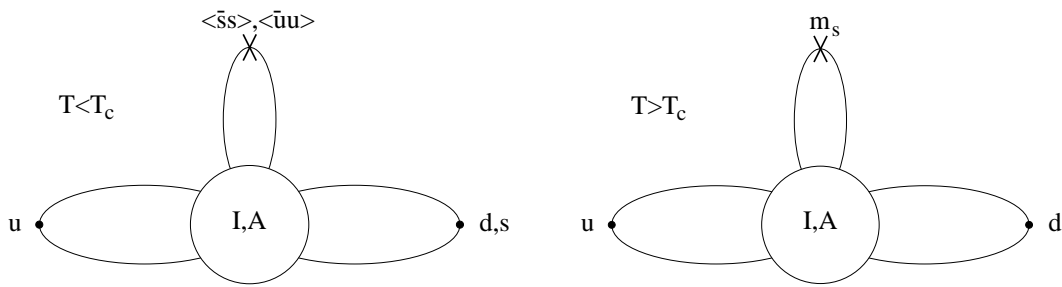}
\caption{\label{fig_anomaly}
Flavor mixing in the $\eta-\eta'$ system below and above the 
chiral phase transition.}
\end{figure}
 
   At temperatures significantly above $T_c$, we expect 
screening to reduce the strength of the $U(1)_A$ violating
interaction. Near $T_c$, screening is not important but
chiral symmetry restoration affects the structure of flavor
mixing in the $\eta-\eta'$ system \cite{Sch_96}, see figure 
\ref{fig_anomaly}. Below $T_c$, there is strong flavor mixing 
between $u,d$ and $s$ quarks. Flavor symmetry is broken,
$\langle\bar uu\rangle \neq\langle\bar ss\rangle$, but the 
$\eta'$ state is almost pure singlet. Above $T_c$, there 
is no flavor mixing between non-strange and strange quarks. As 
a result, the eigenstates in the $\eta-\eta'$ system are the 
non-strange and strange eta components $\eta_{NS}$ and $\eta_S$. 
The anomaly acts only on the non-strange $\eta_{NS}$, so the 
strange $\eta_S$ can become light. This effect is of 
phenomenological interest, because it might enhance 
strangeness production in heavy ion collisions. 
 
\section{Summary}

  There is substantial evidence that non-perturbative effects
are important in QCD, even above the phase transition. This 
evidence comes from an analysis of lattice results for the 
equation of state, the spectrum of screening masses and 
temporal correlation functions above $T_c$. This suggest
that in order to understand the transition region we 
need a more detailed picture of the transition itself.

  We have shown that the chiral phase transition can be
understood as a transition from a disordered instanton
liquid to a correlated phase of instanton-antiinstanton 
molecules. This picture is consistent with both the observation
that not all of the gluon condensate is removed across
the phase transition and with the observed spectrum of
screening masses. It provides interesting predictions
for the behavior of hadronic modes near $T_c$. For 
example, we suggest that $(\pi,\sigma)$-like modes survive
the phase transition and that the structure of flavor
mixing in the $\eta-\eta'$ sector changes near $T_c$.

\section{Acknowledgements}

  The material presented here is based on work done in collaboration 
with E. Shuryak and J. Verbaarschot. I would also like to acknowledge 
many discussions with G. E. Brown, V. Koch and R. Venugopalan.

\end{document}